# Environmental Sensing by Wearable Device for Indoor Activity and Location Estimation

Ming Jin, Han Zou, Kevin Weekly, Ruoxi Jia, Alexandre M. Bayen, and Costas J. Spanos

Department of Electrical Engineering and Computer Sciences,
University of California, Berkeley, CA 94720, USA
{jinming, hanzou, kweekly, ruoxijia, bayen, spanos}@eecs.berkeley.edu

*Abstract*—We present results from a set of experiments in this pilot study to investigate the causal influence of user activity on various environmental parameters monitored by occupant-carried multi-purpose sensors. Hypotheses with respect to each type of measurements are verified, including temperature, humidity, and light level collected during eight typical activities: sitting in lab / cubicle, indoor walking / running, resting after physical activity, climbing stairs, taking elevators, and outdoor walking. Our main contribution is the development of features for activity and location recognition based on environmental measurements, which exploit location- and activity-specific characteristics and capture the trends resulted from the underlying physiological process. The features are statistically shown to have good separability and are also information-rich. Fusing environmental sensing together with acceleration is shown to achieve classification accuracy as high as 99.13%. For building applications, this study motivates a sensor fusion paradigm for learning individualized activity, location, and environmental preferences for energy management and user comfort.

## I. INTRODUCTION

Building intelligence encompasses its ability to sense and understand the activities of occupants to interact with them and achieve goals like comfort and energy efficiency. Individuals perform various activities inside the building. This information, when made available to the building automation and control system, can be very useful. For example, the PMV model proposed by Fanger and adopted by ASHRAE as the primary standard for thermal comfort takes occupant metabolic rate as the most important factor, but it has been widely regarded as the most difficult parameter to measure. Traditional practices model it as constant, which essentially rules it to be irrelevant. By tracking user activities inside the building, it becomes much easier to estimate metabolic rate and improve thermal comfort of the HVAC system.

The activities people are engaged in can be categorized in multiple dimensions, such as movement, time, and location. Typical activities are working in the cubicle or lab area, walking to a meeting, and climbing the stairs. Activities are further characterized by being stationary or moving inside the building. Clearly it is desirable to explore the direction of addressing multiple dimensions of indoor activities for occupancy modeling at a minimal labor and cost.

Wearable electronics, such as the Fitbit and eWatch, are becoming more ubiquitous and carrying more sensors, owing to advances in miniaturization and manufacturing. Currently, technology is at the point where these smart watches can carry a suite of environmental sensors and continuously monitor the wearer's local environment. In the near future, manufacturing techniques will enable the vision of Smart dust: fully-integrated cubic-millimeter size wireless sensors which could be non-intrusively integrated into jewelry, clothes, and implants. In the project we adapted our environmental sensing platform [1] into a watch to conduct wearable sensor studies. The goal in developing the environmental sensing platform was not to be smaller than the current offerings, rather to be small enough to enable these studies.

Our hypothesis is that the environmental measurements made by the wearable infrastructure provide information about where the user is and how he/she moves physically. There are two main contributions by the study. First, we present experimental results that motivate further study using more sophisticated mathematical models, and integrating these types of measurements. We also explore features with good discrimination for the estimation of occupancy activity. The advantage of the proposed method is to leverage the existing sensing infrastructure to fill in the missing information of the tracked activity at minimal additional cost.

This paper is organized as follows. In Section II, we review previous works in individual activity tracking and make a distinction of the current study. The hardware and system infrastructure deployed in the experiment is described in Section III. Section IV focuses on the experimental procedures with detailed setup and data collection. The experimental results and analysis is discussed in Section V. The environmental sensing is used in various classification methods to demonstrate its advantage in Section VI. The conclusion is drawn in Section VII in addition to the suggestion of future work.

## II. RELATED WORK

Activity is a multidimensional concept that is intimately related to the body movement or gestures and the context where it happens. Related works in body motion recognition extensively use miniaturized inertial sensors, namely accelerometers and gyroscopes.

This research is funded by the Republic of Singapore's National Research Foundation through a grant to the Berkeley Education Alliance for Research in Singapore (BEARS) for the Singapore-Berkeley Building Efficiency and Sustainability in the Tropics (SinBerBEST) Program. BEARS has been established by the University of California, Berkeley as a center for intellectual excellence in research and education in Singapore.

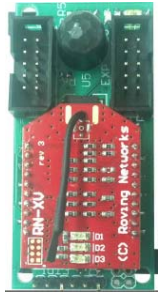

Fig. 1. The hardware design of our environmental sensor

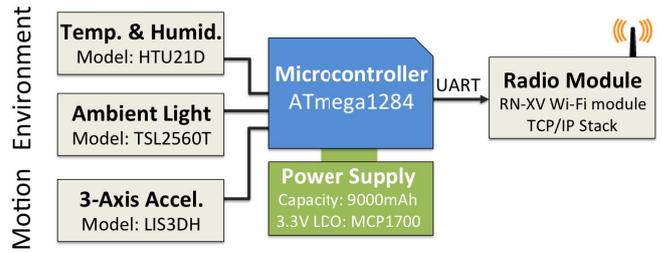

Fig. 2. The physical configuration of our environmental sensor

Single [2] or multiple [3-5] sensors attached to the body can be used to recognize a variety of daily activities with high accuracy. Smart phones such as the iPhone [7], Android phones [8], and Nokia N95 [9] have also been deployed for recognition of common physical activities, such as sitting, standing, walking, running and bicycling. An array of algorithms, mostly implemented by the Weka machine learning toolkit [11], have been deployed in [2-10], such as decision trees, K-nearest neighbors, SVM, Naïve Bayes, Hidden Markov Model, Gaussian Mixture model, and meta-level classifiers categorized by voting used in bagging and boosting, stacking , and cascading.

One promising line of research is the sensor fusion approach, where a number of heterogeneous parameters are collected to infer motion and context. For example, the energy efficient mobile sensing system (EEMSS) [10] uses Wifi, GPS, and microphone together with acceleration to provide information about location (home or outdoor) and background condition (loud or quiet), which can achieve approximately 92.5% detection accuracy. Other types of fusion use vision data [12], magnetometers [13], ultrasonic sensors [14], and active badge [15] in combination with inertial sensors to identify physical motions. These works can leverage the sensing ability of fused components to acquire additional knowledge useful to their application areas, such as home-based rehabilitation, assisted cognition systems, and activity recognition based support of assembly and maintenance tasks.

The present study is different from existing works in the following ways. First, the proposed sensor platform measures a variety of environmental and body movement parameters, such as tri-axial acceleration, temperature, humidity, and light level, enabling us to study the use of environmental sensing for location and activity estimation. Also we focus on indoor activity recognition for the benefits of building energy efficiency and occupant comfort and productivity, as we view the awareness of occupant activity as an essential component of building intelligence.

### III. Hardware and System Infrastructure

We developed a low-cost, battery-powered indoor environmental sensor as shown in Fig. 1. The device collects a rich set of environmental variables, such as temperature, humidity, ambient light, orientation sensing and motion detection. Furthermore, the small size of the sensor makes it possible to be installed easily in any indoor environment. The environmental measurements are delivered and transmitted to centralized servers reliably by leveraging Wifi technology.

The object of our hardware design for the environmental sensor is to provide a permanent, reliable, and easily deployable sensing infrastructure that can operate for years inside a smart building. Moreover, the sensor should be wearable and portable for occupants to carry.

Fig. 2 demonstrates the physical configuration of our environmental sensor. The logical center of the design is Atmel's ATmega1284 microcontroller. The sensor (which was not inititally designed to be wearable) is powered via a 3.7-volt Lithium-Thionyl Chloride battery with a nominal battery life of 9Ah. We chose Microchip's RN-XV WiFi module as the communication module since it has been proved to provide the most reliable communications in real deployments. Measuring Temperature and Humidity from the local environment is accomplished by Measurement Specialties HTU21D transducer. The stated accuracy of the humidity sensor is ±2%RH typical over the 20%RH to 80%RH range, and up to ±3%RH outside of this range. The stated accuracy of the temperature sensor is typically ±0.3℃ and maximally ±0.4℃ over the range of roughly 5℃ to 60℃. Ambient visible light is measured by AMS TSL2560 sensor, including two light-sensing photodiodes: one measures visible and IR light from 300 nm to 1100 nm, and the other measures IR light from 500 nm to 1100 nm. The LIS3DH Accelerometer by ST Microelectronics is chosen to measure the orientation and acceleration variation of the system.

The network configuration of our environmental sensor is illustrated in Fig. 3. All the environmental sensors and personal watches are connected to centroid servers using WiFi. As shown in Fig. 3, all the environmental measurements are sent to a local server and an Internet server simultaneously. All the data are stored on an on-board PostgreSQL server which also contains metadata about the incoming measurements. The internet server is similar to the local server. It uses an online cloud database to store all of the data.

### IV. Experimental Design

Our present study is carried out in the Center for Research in Energy Systems Transformation (CREST) located in Cory Hall on UC Berkeley campus, which is an office with cubicles and a lab area. We also study several other places inside and outside near Cory.

We identify the following eight (8) activities of interests: sitting in the cubicle / lab, walking indoor / outdoor, running indoor, taking an elevator, climbing stairs, and resting. While the physical motions (sitting, standing, walking, running, and climbing) constitute a standard set of training, the associated locations reflect the characteristics of indoor application. In the CREST testbed, the lab and cubicle are in separately controlled environment zones, For instance, knowing in advance that the occupant has just walked from outdoor to indoor, and decided to climbing the stairs is useful in cooling environment upon arrival. We describe the physical setup and the procedure as follows.

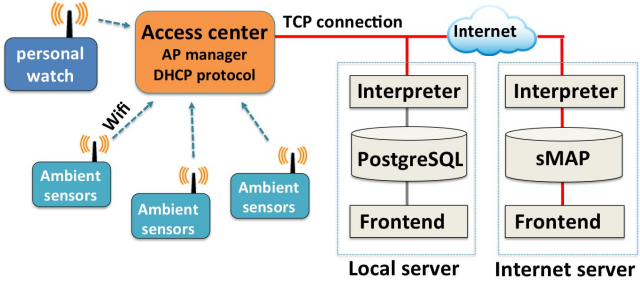

Fig. 3. The network configuration of the wearable sensor and ambient sensors

*A. Physical Configuration*

The primary location of the body sensor placement is the wrist to feel like a watch or wristband. Commercial products like FitBit and eWatch also choose the wrist for its convenience and to enhance user experience. We also benefit from being able to measure light level and temperature by placing it in exposure, which is reasonable for watch.

We also deploy sensors in the lab and the cubicle to collect temperature, humidity, and light level measurements. Our hypothesis is that the correlation between wearable and environmental sensing provides location estimation. Since we can distinguish outdoor and indoor locations by the watch measurements alone, and the environment in the stairs is mostly invariant, we did not place any sensors in these areas. Lastly, all the sensors are properly calibrated before deployment.

*B. Experimental Procedure*

After the experimental configuration, the subject performs each of the eight activities individually, recording the time, and labeling it with the ground truth. We also ask the subjects to act normally and stick to the daily routines. For instance, the samples of indoor walking are taken at different times during the daily works, rather than a long period of walking just to collect the data.

The above procedure is carried out by two subjects independently. They have different activity patterns to stand the test of generalizability of the experiments. For example, individual A works mostly in the lab while B spends most of the time in the cubicle. The total time of measurements for each activity is shown in Table I.

## V. RESULTS AND ANALYSIS

Our main hypothesis for the study is that the environmental measurements collected by the wearable device can provide useful information about the current location and motion of the occupants. The subsequent hypotheses are formed around the types of measurements, including temperature, humidity and light level. Following the observation of the experiments, we

TABLE I. SUMMARY OF EXPERIMENTAL MEASUREMENTS

| | Climb stairs | Take elev. | Walk out. | Walk indr. | Run indr. | Sit lab | Sit cub. | Rest |
|---|---|---|---|---|---|---|---|---|
| Time (min) | 44 | 25 | 41 | 37 | 52 | 176 | 169 | 26 |

a. The amount of data valid after preprocessing rounded to the nearest minutes

propose a set of features to capture the phenomena, and conduct statistical testing to formally verify the hypothesis.

***Hypothesis I:*** *When the user starts to walk or run, then the temperature reading will drop as is reflected by the negative gradients and the larger standard deviation.*

Temperature generally reads higher on a wearable sensor as compared with environment as influenced by the metabolic heat generation of the occupant. Also, we observed that it is obviously lower during dynamic activities such as walking and running as compared to stationary states like sitting and standing.

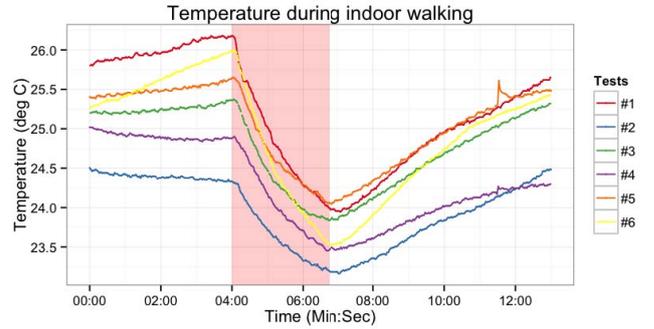

Fig. 4. Temperature decreases during walking due to the cooling effect

This is likely to be caused by the cooling effects from increased airflow over the watch during walking (i.e. user walks with a normal gait, including swinging the arms), as shown in Fig. 4.

We use temperature gradients and standard deviation defined below to capture this trend:

$$\hat{\nabla} Tp_k = \frac{1}{\lfloor w/2 \rfloor + 1} \sum_{i=0}^{\lfloor w/2 \rfloor} \frac{Tp_{k-i} - Tp_{k-\lfloor w/2 \rfloor - i}}{T_{k-i} - T_{k-\lfloor w/2 \rfloor - i}} \quad (1)$$

$$sd(Tp)_k = \sqrt{\frac{1}{w} \sum_{i=0}^{w-1} \left(Tp_{k-i} - \overline{Tp_{k-w+1:k}}\right)^2} \quad (2)$$

where $w$ is the window length, $Tp_k$ and $T_k$ are the temperature and time at index $k$ in the time series, and the mean temperature of the segment indexed by $k-w+1:k$ is denoted as $\overline{Tp_{k-w+1:k}}$.

Fig. 5 and 6 show the conditional distribution of each feature given the set of stationary activities (sitting in lab / cubicle, standing in elevator) and the set of dynamic activities (walking / running indoor). The temperature gradient and standard deviation have good separability for static / dynamic activities. That is, the temperature gradient tends to be negative and standard deviation tends to be larger for dynamic activities.

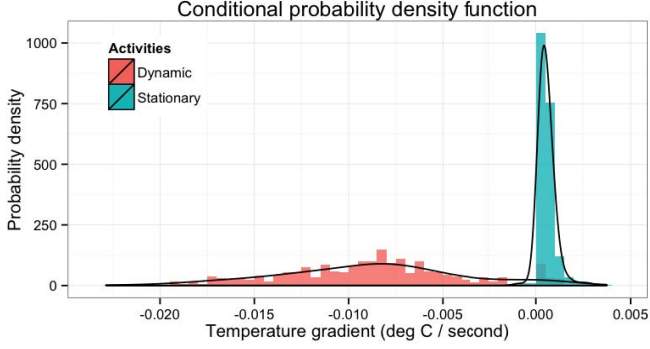

Fig. 5. Probability density and histograms of temperature gradients for stationary (green) and dynamic (red) activities[1]

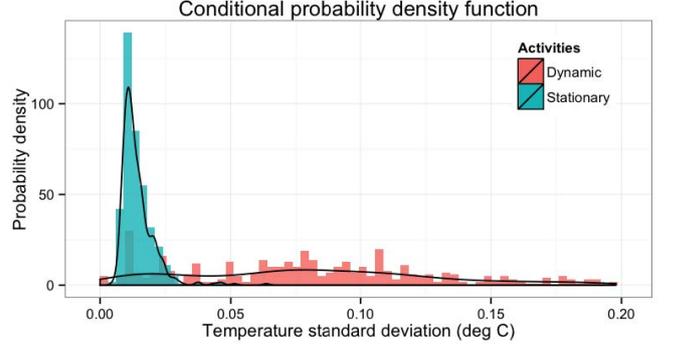

Fig. 6. Probability density and histograms of temperature standard deviation for stationary (green) and dynamic (red) activities

The feature separability can be measured by the Jensen-Shannon divergence (JSD), also known as the information radius or total divergence to the average, which is given by:

$$JSD(P\|Q) = \frac{1}{2}D(P\|M) + \frac{1}{2}D(Q\|M) \quad (3)$$

where $M = \frac{1}{2}(P+Q)$ is the mixture distribution, and the Kullback-Leibler (KL) divergence, $D(P\|Q)$ is defined by:

$$D(P\|M) = \int p(x)\ln\frac{p(x)}{q(x)}dx \quad (4)$$

The Jensen-Shannon divergence can be considered as a symmetrized and smoothed version of the KL divergence an can be shown to have upper bound of $\ln(2)$. The larger the JSD is, the better the separability the feature has achieved. JSD of value .5911 is obtained for the temperature gradient and standard deviations, which are listed in Table II.

We apply the classical permutation test to verify the statistical significance of the observations. The idea is to randomly permute the labels of activities and each time obtain a JSD. The null hypothesis is that the observed JSD between dynamic and stationary activities for temperature gradients, or humidity standard deviation, is identical to the JSD of data with the randomly permuted labels of stationary and dynamic activities, namely:

$$H_0^{\nabla Tp}: JSD_{Observed}^{\nabla Tp} = JSD_{Permuted}^{\nabla Tp} \quad (5)$$

$$H_0^{sd(Tp)}: JSD_{Observed}^{sd(Tp)} = JSD_{Permuted}^{sd(Tp)} \quad (6)$$

We can then reject the null hypothesis if the observed JSD significantly deviates from the mean of the JSD distribution in the permutation test. The permutation test does not require normal distribution of data, and obtains the distribution by bootstrapping the samples, which makes it widely adopted as a non-parametric testing procedure.

The result of the permutation test is shown in Table II. As can be seen, the observed JSD is significantly higher than the distribution obtained in the permutation test. The *p*-value of .001 indicates the probability of observing the JSD as significant as in the experiment under null hypothesis, which is reliable enough to reject the null. It is, therefore, concluded that the features based on temperature gradients and standard deviations can achieve high separability for stationary and dynamic activities.

***Hypothesis II:*** *When the user rests after a period of walking or running, the humidity will increase and exhibit high variations.*

Sweating is an important mechanism for keeping the body cool during or after exercise by removing excess heat through evaporation. The change in humidity above the skin might be unnoticeable to human, but is repeatedly picked up by the wearable sensor, as is shown in Fig. 7. There is a significant increase in variation after a period of exercise (walking or running) than during the exercise. This is likely to be caused by the airflow over the sensor that quickly takes the mist away during motion, while the mist gathered to form a boundary layer around the body when the user stays still. Hence we use humidity standard deviation defined below to differentiate resting after exercise from other activities:

$$sd(Hm)_k = \sqrt{\frac{1}{w}\sum_{i=0}^{w-1}\left(Hm_{k-i} - \overline{Hm_{k-w+1:k}}\right)^2} \quad (7)$$

Evidence of this effect is seen by Fig. 8, which demonstrates separation between resting with other activities. The null hypothesis for the permutation test is given by:

$$H_0^{sd(Hm)}: JSD_{Observed}^{sd(Hm)} = JSD_{Permuted}^{sd(Hm)} \quad (8)$$

The test statistics given in Table II shows that the separation is statistically significant with *p*-value .001.

***Hypothesis III:*** *When the user sits in a lab/cubicle (with different lighting conditions), then the light level will reflect the characteristics of that location as compared to the reference points.*

We observed that different locations in the building exhibit different light levels, while the light level for the same place during a short period of time tends to be stable, as shown in Fig. 9, which plots the light level for cubicle, lab and user watch.

---

[1] For all the conditional probability plots, the features are scaled by a constant in order to fit the probability density and histograms in the same graph. The histograms are also bootstrapped for estimating the empirical distribution.

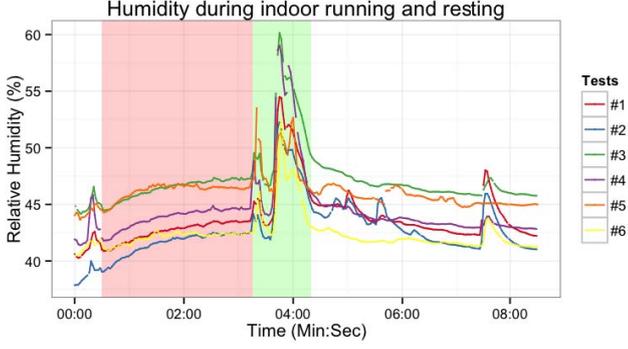

Fig. 7. Humidity exhibits high variation (green) after physical activity (red)

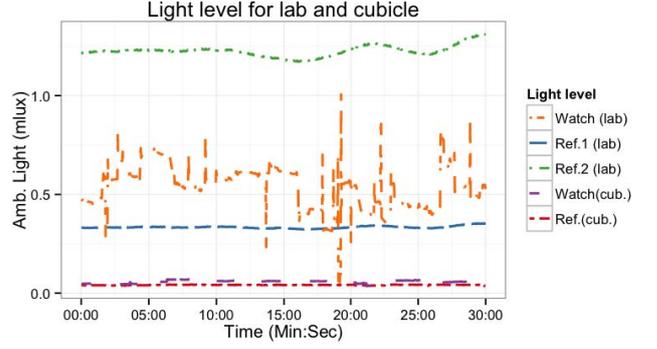

Fig. 9. Light level in lab and cubicle as compared to the wearable sensor

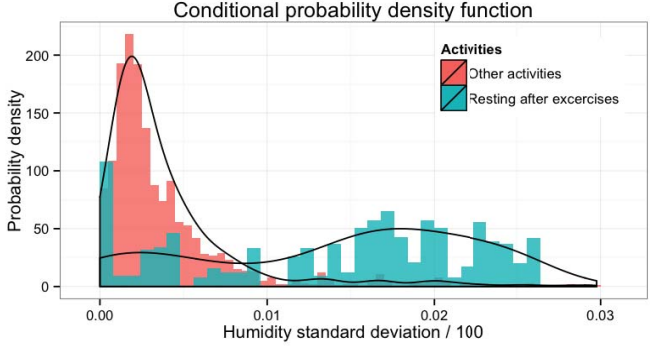

Fig. 8. Humidity standard deviation for resting after physical activity (green), and other activities (walking/running indoor, sitting in lab/cubicle and climbing)

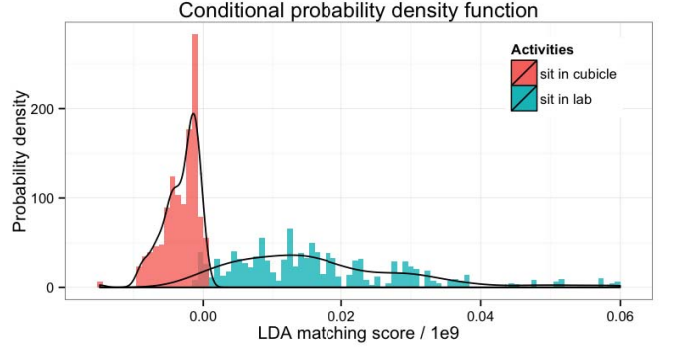

Fig. 10. LDA matching scores (rescaled by 1e9 for graphical representation) calculated for sitting in lab (red) and lab (green)

We consider the light level conditioned on different locations, i.e. lab/cubicle, to be independent identically distributed (i.i.d.) Gaussian random variables with different mean values and variances. Sample mean and sample variance of readings from reference sensors are used to estimate the light distributions in lab/cubicle. We apply linear discriminant analysis (LDA) to determine the location based on a series of light sensor readings. The idea is to compare the probability of given observations conditioned on different locations. We use the log of the likelihood ratio to indicate the location:

$$LR = \sum_{i=1}^{N}\left[\ln(\frac{\sigma_{cub}^2}{\sigma_{cub}^2}) + \frac{(x_i - \mu_{cub})^2}{2\sigma_{cub}^2} - \frac{(x_i - \mu_{lab})^2}{2\sigma_{lab}^2}\right] \quad (9)$$

where $\mu_{cub}$, $\mu_{lab}$, $\sigma_{cub}^2$ and $\sigma_{lab}^2$ are sample means and sample variances of reference sensor readings placed in the cubical and lab, $x_i$ is one light level observation from a series of readings, $N$ is the duration of observations. If $LR$ is below some threshold, we predict the observations are taken from a cubicle, otherwise from a lab.

The separation of $LR$ of light level observations in a lab/cubicle is illustrated in Fig. 10. We can set zero as the threshold to differentiate different locations. The null hypothesis for the permutation test is given by:

$$H_0^{LR} : JSD_{\text{Observed}}^{LR} = JSD_{\text{Permuted}}^{LR} \quad (10)$$

which reveals $p$-value as significant as .001, a clear indication of good feature separability.

***Hypothesis IV:*** *When the user is climbing the stairs, then the light level will exhibit periodic pattern and the magnitude will reflect the floor he/she has reached.*

As stated before, different locations present different light levels. When the user is climbing the stairs, the light level will change periodically. This is because stairs is composed of different light zones that appear "periodically". Fig. 11 illustrates the light changes when the user is climbing the stairs. The light is dimmer in the intermediate landing and brighter in the main floor levels. We also indicated the light level as measured by our reference points for each floor which exhibits noticeable difference.

We can define a template light pattern of climbing stairs by exploiting the reference light levels at different main floors and the average velocity of climbing stairs. In practice, users might climb stairs at different rates and ends climbing at any floor. Therefore, the "period" and "duration" of the light level time series might vary randomly. Considering the scenarios aforementioned, we apply dynamic time warping (DTW) to measure similarity between the observed temporal sequences with the template light pattern. DTW is an algorithm that calculates an optimal match between two sequences by arranging all sequence points and thus can allow acceleration of deceleration of signals along the time dimension. The distance formula of DTW is given by:

$$DTW[i,j] = (s[i] - t[j])^2 + \min\{DTW[i-1,j], DTW[i,j-1], DTW[i-1,j-1]\} \quad (11)$$

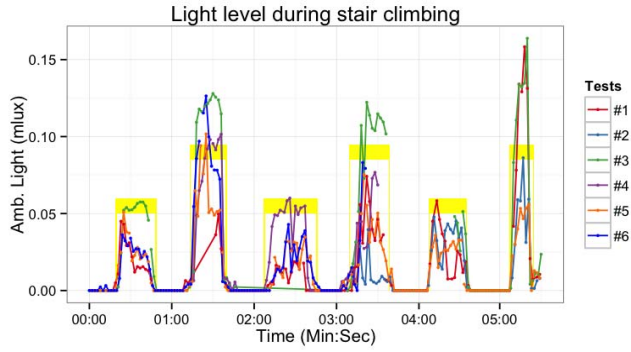

Fig. 11. Periodic pattern as observed during climbing the stairs

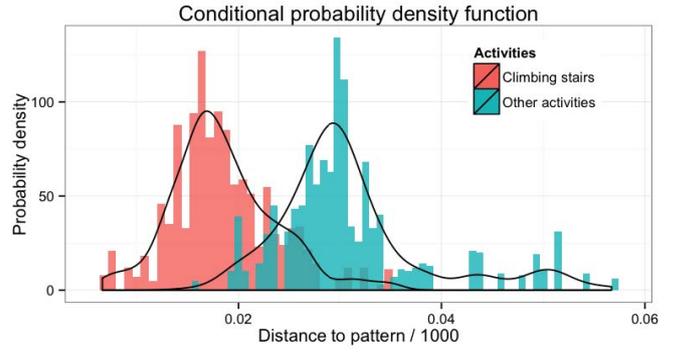

Fig. 12. DTW distance to the periodic pattern of light level in stairs for stair climbing (red) and other activities (green)

where $s[i]$ is the observed light readings at time $i$ and $t[j]$ is the template light level sequences at time $j$. As is shown in Fig. 12, the distances measured by DTW can be used to determine if the user is climbing stairs. We examine the null hypothesis given below for the permutation test:

$$H_0^{DTW}: JSD_{Observed}^{DTW} = JSD_{Permuted}^{DTW} \quad (12)$$

which reveals statistical significance of $p$-value as .001.

TABLE II. SUMMARY OF STATISTICS FROM PERMUTATION TEST

| | $JSD_{Obv.}$ | $\mu(JSD_{Perm.})$ | $sd(JSD_{Perm.})$ | $p$-value[a] |
|---|---|---|---|---|
| $H_0^{\nabla Tp}$ | .5911 | .0334 | .0045 | 1e-3* |
| $H_0^{sd(Tp)}$ | .4664 | .0216 | .0033 | 1e-3* |
| $H_0^{sd(Hm)}$ | .1954 | .0181 | .0046 | 1e-3* |
| $H_0^{LR}$ | .5407 | .0191 | .0047 | 1e-3* |
| $H_0^{DTW}$ | .3995 | .0062 | .0036 | 1e-3* |

[b.] p-value is calculated by the rank of observation in the sequence divided by the total number of tests

## VI. SENSOR FUSION FOR ACTIVITY RECOGNITION

Sensor fusion is an approach of combination, association and correlation of heterogeneous sensor measurements to make better decision. Various approaches have been proposed for information combination [17]. In this study we deployed a list of common classifiers as supported by the Weka machine learning toolkit [11], as shown in Table III, to illustrate the advantage of including environmental sensing for activity recognition. For the comparison, we did not pursue sophisticated time-series analysis such as counting steps. We use the standard deviation of the x,y,z-axis for the acceleration data. The environment data is also based on gradients and standard deviations as discussed in the previous section.

From Table III, we can see that environmental sensing alone can achieve activity classification accuracy as high as 97.42% as compared with 54.19% with only acceleration data. If we fuse both types of measurements, the classification accuracy is 99.13% with Random Forests. All the accuracy is calculated by the standard 10-fold cross-validation. We attribute this improvement to the richer set of information that environmental sensing possesses, which not only distinguishes dynamic from static activities, also it can reliably estimate the

TABLE III. ACCURACY OF CLASSIFIERS W/O ENVIRONMENTAL SENSING

| | Acceleration | Environment | Acc. + Environ. |
|---|---|---|---|
| Naïve Bayes | .5309 | .8706 | **.9459** |
| Bayes Net | .5396 | .9372 | **.9843** |
| Logistic regression | .5194 | .8995 | **.9728** |
| Multi-layer Perception | .5099 | .8612 | **.9379** |
| RBF Network | .4220 | **.5623** | .5364 |
| Decision Table | .5419 | .9023 | **.9226** |
| J48 Tree | .5403 | .9599 | **.9854** |
| Random Tree | .5033 | .9616 | **.9850** |
| Random Forrest | .5033 | .9742 | **.9913** |

[c.] All the algorithms are implemented in Weka machine learning toolkit.

TABLE IV. REPRESENTATIVE CONFUSION MATRIX FOR DECISION TABLE WITH ONLY ACCELERATION MEASUREMENTS FOR CLASSIFICATION

| Activity | Classified As | | | | | | | |
|---|---|---|---|---|---|---|---|---|
| | a | b | c | d | e | f | g | h |
| a: climb | 50 | 0 | 0 | 19 | 109 | 23 | 0 | 0 |
| b: elevator | 0 | 0 | 0 | 0 | 0 | 0 | 89 | 24 |
| c: rest | 0 | 0 | 16 | 0 | 6 | 19 | 27 | 1 |
| d: run/ind | 1 | 0 | 8 | 199 | 0 | 0 | 6 | 24 |
| e: walk/ind. | 29 | 0 | 1 | 0 | 152 | 103 | 2 | 0 |
| f: walk/out. | 22 | 0 | 0 | 0 | 94 | 146 | 3 | 0 |
| g: sit/lab | 0 | 0 | 10 | 0 | 0 | 17 | 276 | 561 |
| h: sit/cub. | 0 | 0 | 0 | 0 | 0 | 0 | 117 | 714 |

TABLE V. REPRESENTATIVE CONFUSION MATRIX FOR RANDOM FORESTS WITH ACCELERATION AND ENVIRONMENTAL MEASUREMENTS

| Activity | Classified As | | | | | | | |
|---|---|---|---|---|---|---|---|---|
| | a | b | c | d | e | f | g | h |
| a: climb | 197 | 0 | 0 | 2 | 2 | 0 | 0 | 0 |
| b: elevator | 0 | 112 | 0 | 0 | 0 | 0 | 0 | 1 |
| c: rest | 1 | 1 | 63 | 0 | 2 | 0 | 0 | 1 |
| d: run/ind | 1 | 0 | 8 | 228 | 0 | 0 | 0 | 1 |
| e: walk/ind. | 3 | 0 | 0 | 0 | 284 | 0 | 0 | 0 |
| f: walk/out. | 1 | 0 | 0 | 0 | 0 | 264 | 0 | 0 |
| g: sit/lab | 0 | 0 | 0 | 0 | 0 | 0 | 864 | 0 |
| h: sit/cub. | 0 | 1 | 0 | 0 | 0 | 0 | 0 | 830 |

location based on the characteristic ambient parameters. This is illustrated in the confusion matrix in Table IV and V, which demonstrates that environmental sensing can successfully distinguish activities in both the motion and location dimensions that would otherwise confuse classification methods based on acceleration only. For instance, activity sets such as sit in lab / cubicle, walk indoor / outdoor and climbing, taking an elevator and stationary activities, all have similar acceleration profiles, but by examining the ambient parameters such as light level, we can reliably make correct predictions. In addition, environmental sensing provides extra information to further distinguish activities with nuances, such as humidity variation to tell apart resting after physical activity and sitting in lab / cubicle.

Future works can use approximate inference, such as particle filtering, to efficiently perform online activity recognition.

## VII. Conclusion

We have shown via experimentation that indoor occupancy activity can be recognized and classified by leveraging the environmental measurements measured by the occupant-carried environmental sensing wearable devices and environmental sensor network. We proposed numerous features for activity and location recognition. We adopted temperature gradients and standard deviation to distinguish dynamic or static activities of occupants. Our statistical testing results have shown that these two features can achieve high reparability for different activities. We also found that the humidity standard deviation can be employed to differentiate resting after physical activities from other activities. Based on our observation, different locations in buildings exhibit different light levels. Therefore, our experimental results have demonstrated that the light level measured by the wearable devices can reflect the characteristics of occupant's location as compared to the reference environmental sensors. Furthermore, another interesting observation is that the light level measured by the wearable device can exhibit periodic pattern and the magnitude can reflect the floor the occupancy has reached when he or she is climbing the stairs.

These features are statistically shown to have good separability and are also information-rich. In addition to these, we have shown that fusing environmental sensing together with acceleration can achieve classification accuracy as high as 99.13%. Actually environmental sensing can successfully distinguish activities that would confuse acceleration under several circumstances.

This paper provides a motivation of a sensor fusion paradigm for learning individualized activity, location, and environmental preferences for energy management and user comfort. Future work can be focused on the information fusion between these environmental related features with other monitoring systems, such as power metering system and WiFi-based indoor positioning system [18], to obtain more accurate, robust and reliable recognition of indoor occupancy activities. We would also develop online algorithms based on environmental sensing to track user activities in real-time to provide useful information to the building management system.